\documentclass[aps,10pt,prl,notitlepage,tightenlines,nofootinbib,twocolumn,superscriptaddress]{revtex4-2}

\usepackage{amsmath}
\usepackage{amssymb,amsthm}
\usepackage{mathtools}
\usepackage{mathrsfs}
\usepackage{relsize}
\usepackage{bm}

\usepackage{epsfig,graphics,graphicx,color,xcolor}
\usepackage{soul} 
\usepackage{cancel} 
\usepackage{slashed}	
\usepackage{subfigure}
\usepackage{array}
\usepackage{multirow}

\usepackage{ragged2e}
\usepackage{lineno}
\usepackage{natbib}
\usepackage{hyperref}
\hypersetup{colorlinks=true,linkcolor=red,anchorcolor=red,citecolor=orange, filecolor=brown,urlcolor=red,bookmarksnumbered=true,
pdfview=FitB
}
\graphicspath{ {image/} }
\usepackage{enumitem}
\usepackage{ulem}

\colorlet{darkgreen}{green!50!black}
\colorlet{brightyellow}{yellow!75!red}
\colorlet{orange}{red!50!yellow}
\colorlet{darkgray}{gray!50!black}

\def\dd{{\mathrm{d}}}

\makeatletter
\newcommand*{\transpose}{%
  {\mathpalette\@transpose{}}%
}
\newcommand*{\@transpose}[2]{%
  \raisebox{\depth}{$\m@th#1\intercal$}%
}
\makeatother 

\begin{document}

\title{Gravitational form factor $D$ of charmonium from shear stress}

\author{Tianyang Hu}
\affiliation{Department of Modern Physics, University of Science and Technology of China, Hefei, Anhui 230026, China}
\affiliation{Institute of Modern Physics, Chinese Academy of Sciences, Lanzhou 730000, China}
\affiliation{School of Nuclear Science and Technology, University of Chinese Academy of Sciences, Beijing 100049, China}

\author{Xianghui Cao}
\affiliation{Department of Modern Physics, University of Science and Technology of China, Hefei, Anhui 230026, China}

\author{Siqi Xu}
\affiliation{Institute of Modern Physics, Chinese Academy of Sciences, Lanzhou 730000, China}
\affiliation{School of Nuclear Science and Technology, University of Chinese Academy of Sciences, Beijing 100049, China}
\affiliation{Department of Physics and Astronomy, Iowa State University, Ames, Iowa 50010, U.S.}

\author{Yang Li}
\thanks{Corresponding author}
\affiliation{Department of Modern Physics, University of Science and Technology of China, Hefei, Anhui 230026, China}
\affiliation{Anhui Center for Fundamental Sciences (Theoretical Physics), University of Science and Technology of China, Hefei, 230026, China}

\author{Xingbo Zhao}
\affiliation{Institute of Modern Physics, Chinese Academy of Sciences, Lanzhou 730000, China}
\affiliation{School of Nuclear Science and Technology, University of Chinese Academy of Sciences, Beijing 100049, China}
\affiliation{CAS Key Laboratory of High Precision Nuclear Spectroscopy, Institute of Modern Physics, Chinese Academy of Sciences, Lanzhou 730000, China}

\author{James P. Vary}
\affiliation{Department of Physics and Astronomy, Iowa State University, Ames, Iowa 50010, U.S.}

\date{\today}

\begin{abstract}
Based on our recent analysis of the hadronic matrix element of the stress-energy tensor in covariant light front dynamics, we extract the charmonium gravitational form factor $D(Q^2)$ from shear stress $T^{12}$. This is in contrast to our recent work using the (light-front) energy density $T^{+-}$. Indeed, by comparing these two currents, we identify terms that are responsible for the violation of the current conservation. Numerical results based on basis light-front quantization show that the violation effects are small and the $D$-term extracted from the two currents are close to each other, hence validating our previous work using $T^{+-}$. 
\end{abstract}

 \maketitle

\paragraph{Introduction} 

There is intense current interest in the hadronic gravitational form factors (GFFs) \cite{Polyakov:2002yz, Cebulla:2007ei, Mai:2012cx, Mai:2012yc, Jung:2013bya, Cantara:2015sna, Hudson:2017oul, Kumano:2017lhr, Yang:2018nqn, Shanahan:2018pib, Shanahan:2018nnv, Burkert:2018bqq, Polyakov:2018exb, Kumericki:2019ddg, Anikin:2019kwi, Ozdem:2019pkg, Varma:2020crx, Chakrabarti:2020kdc, Pefkou:2021fni, Burkert:2021ith, Dutrieux:2021nlz, Metz:2021lqv, More:2021stk, Duran:2022xag, Tong:2022zax, Mamo:2022eui, Fujita:2022jus, Xing:2022mvk, Ozdem:2022zig, Freese:2022jlu, More:2023pcy, Pasquini:2023aaf, GarciaMartin-Caro:2023klo, Chakrabarti:2023djs, Amor-Quiroz:2023rke, Guo:2023pqw, Kou:2023azd, Dehghan:2023ytx, Cao:2023ohj, Won:2023zmf, Krutov:2023ztx, Li:2023izn, Amor-Quiroz:2023rke, Alharazin:2023uhr, Hackett:2023rif, Hackett:2023nkr, Li:2024vgv, Hagiwara:2024wqz, Nair:2024fit, Fujii:2024rqd, Xu:2024cfa, Cao:2024rul, Cao:2024fto, Sultan:2024hep, Yao:2024ixu, Wang:2024fjt, Cao:2024zlf, Panteleeva:2024abz, Broniowski:2024mpw, Panteleeva:2024zya, Freese:2024rkr, Lorce:2025oot, Klest:2025rek, Guo:2025jiz, Goharipour:2025lep}. These quantities parametrize the hadronic stress-energy tensor $T^{\mu\nu}$ \cite{Pagels:1966zza, Kobzarev:1962wt}, and are related to the energy and stress distributions within the strong interaction bound states \cite{Teryaev:2016edw, Polyakov:2018zvc, Burkert:2023atx, Lorce:2025oot}. 
Direct experimental measurement of GFFs is not possible at this point, due to the feeble nature of gravity on the hadronic level. Present experimental access is based on the connection between GFFs and the generalized parton distributions (GPDs) \cite{Polyakov:2002yz, Ji:1996nm}. In particular, GFFs of the pion and the proton are extracted from electron-hadron scattering including its crossing processes \cite{Kumano:2017lhr, Burkert:2018bqq, Burkert:2021ith, Kumericki:2019ddg, Dutrieux:2021nlz, Duran:2022xag, Klest:2025rek}. In this frontier research area, the forthcoming electron-ion colliders are expected to play an important role to investigate these quantities \cite{Accardi:2012qut, Anderle:2021wcy, Accardi:2023chb}. 

Despite of these advances, major puzzles associated with GFFs exist. For example,  the $D$-term still lacks a clear physical interpretation, and its relation to the mechanical stability of the hadron is not clear \cite{Perevalova:2016dln, Polyakov:2018zvc, Ji:2021mfb, Freese:2022jlu, Li:2024vgv, Freese:2024rkr, Lorce:2025oot}. In this regard, the Hamiltonian formulation, with the associated wave function representation, is poised to provide a direct microscopic interpretation of the GFFs \cite{Brodsky:2000ii, Cao:2023ohj, Cao:2024rul, Cao:2024fto}. 
This is demonstrated in our recent work on the computation of GFFs of charmonium \cite{Xu:2024cfa}. In Refs.~\cite{Cao:2023ohj}, we show that the $D$-term can be extracted from the light-front energy (i.e. Hamiltonian) density $T^{+-}$, where the light-cone coordinates are defined as $v^\pm = v^0 \pm v^3$, $\vec v_\perp = (v^1, v^2)$ for a 4-vector $v^\mu$. This operator was shown to be properly renormalized, and is not affected by spurious terms caused by the violation of the Poincaré symmetry in the non-perturbative regime \cite{Cao:2023ohj}. For charmonium, we constructed the light-front Hamiltonian density operator $T^{+-}$ by localizing the phenomenological light-front Hamiltonian operator $P^-$ on the transverse plane with an impulse ansatz \cite{Xu:2024cfa}. The obtained operator  $T^{+-}$ is then used to compute the GFF $D$. 

In a recent work \cite{Cao:2024rul}, we further investigated the current components of the light-front stress-energy tensor $T^{\mu\nu}$ systematically in covariant light-front dynamics \cite{Close:1971bp, Osborn:1972dy, Berestetsky:1977zk, Karmanov:1991fv, Karmanov:1996un, Carbonell:1998rj, Carbonell:1999pt, Carbonell:2000rr, Karmanov:2002qu, Brodsky:2003pw}, and identified $T^{++}_i$, $T^{+a}_i$, $T^{+-}_i$ and $T^{12}_i$ as ``good currents" that are not contaminated by the spurious terms. Here, $a=1,2$ and $i$ enumerates the constituents of the system. Thus, these good currents are free of uncanceled divergences.  They can be used to extract the full set of GFFs, $A_i(Q^2)$, $D_i(Q^2)$ and $\bar c_i(Q^2)$, which allows a decomposition of the hadronic energy and stress in terms of the constituents, e.g. quarks and gluons. For the total GFFs, current conservation requires $\bar c(Q^2) \equiv \sum_i \bar c_i(Q^2) = 0$. As such, the GFF $D$ can be extracted from both the light-front energy density $T^{+-}$ and the shear stress $T^{12}$. In this work, we investigate the GFF $D$-term of charmonium extracted from the shear stress $T^{12}$, and compare it with our previous result extracted from the light-front energy density $T^{+-}$. 
The advantage is that it does not depend on the impulse ansatz \cite{Xu:2024cfa}, nor does it depend on the current conservation  $\partial_\mu T^{\mu\nu} = 0$. 

\begin{figure}
\centering
\subfigure[\ \label{fig:etac_1S_T12vsT+-}]{\includegraphics[width=0.4\textwidth]{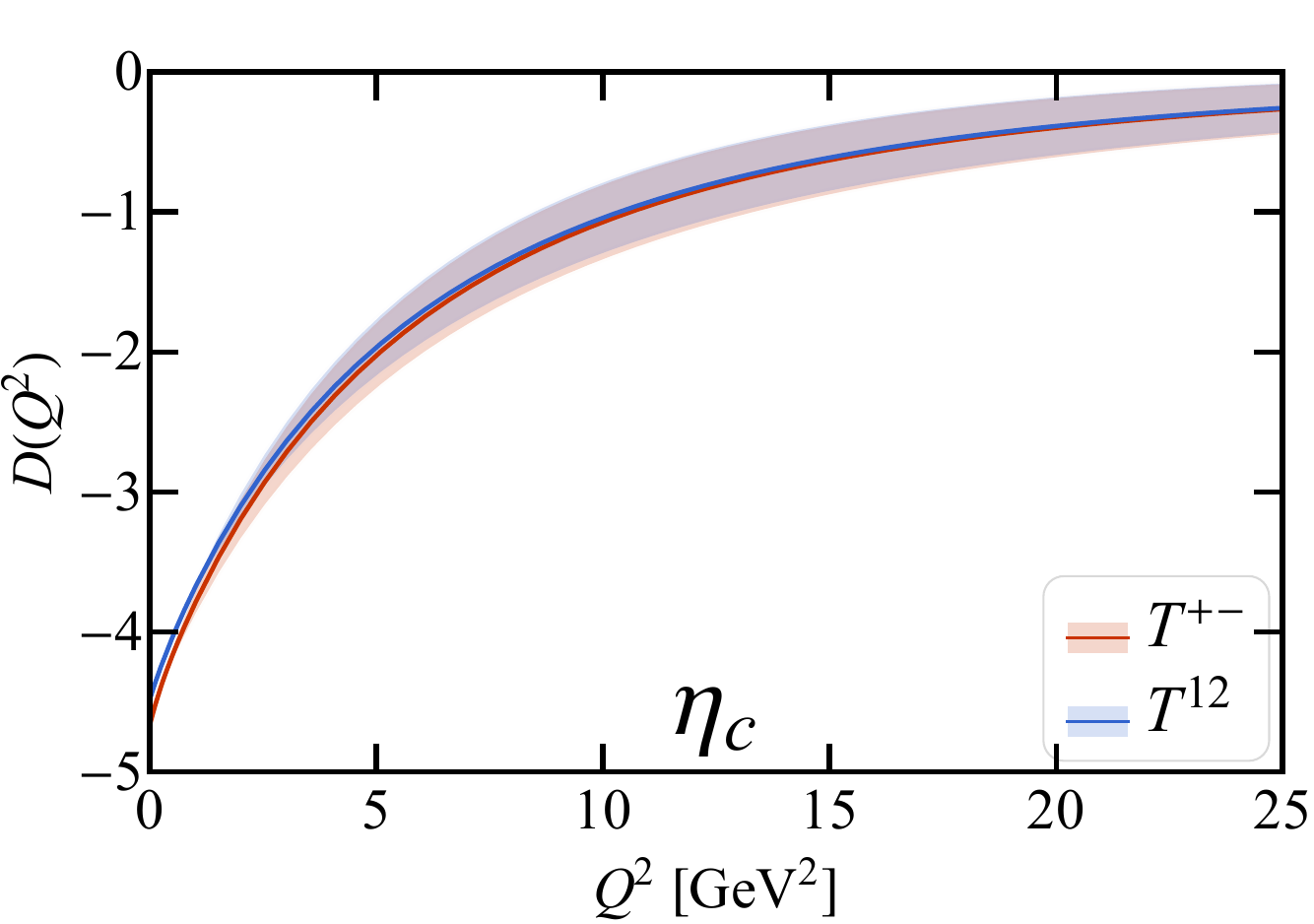}}
\subfigure[\ \label{fig:chic0_1P_T12vsT+-}]{\includegraphics[width=0.4\textwidth]{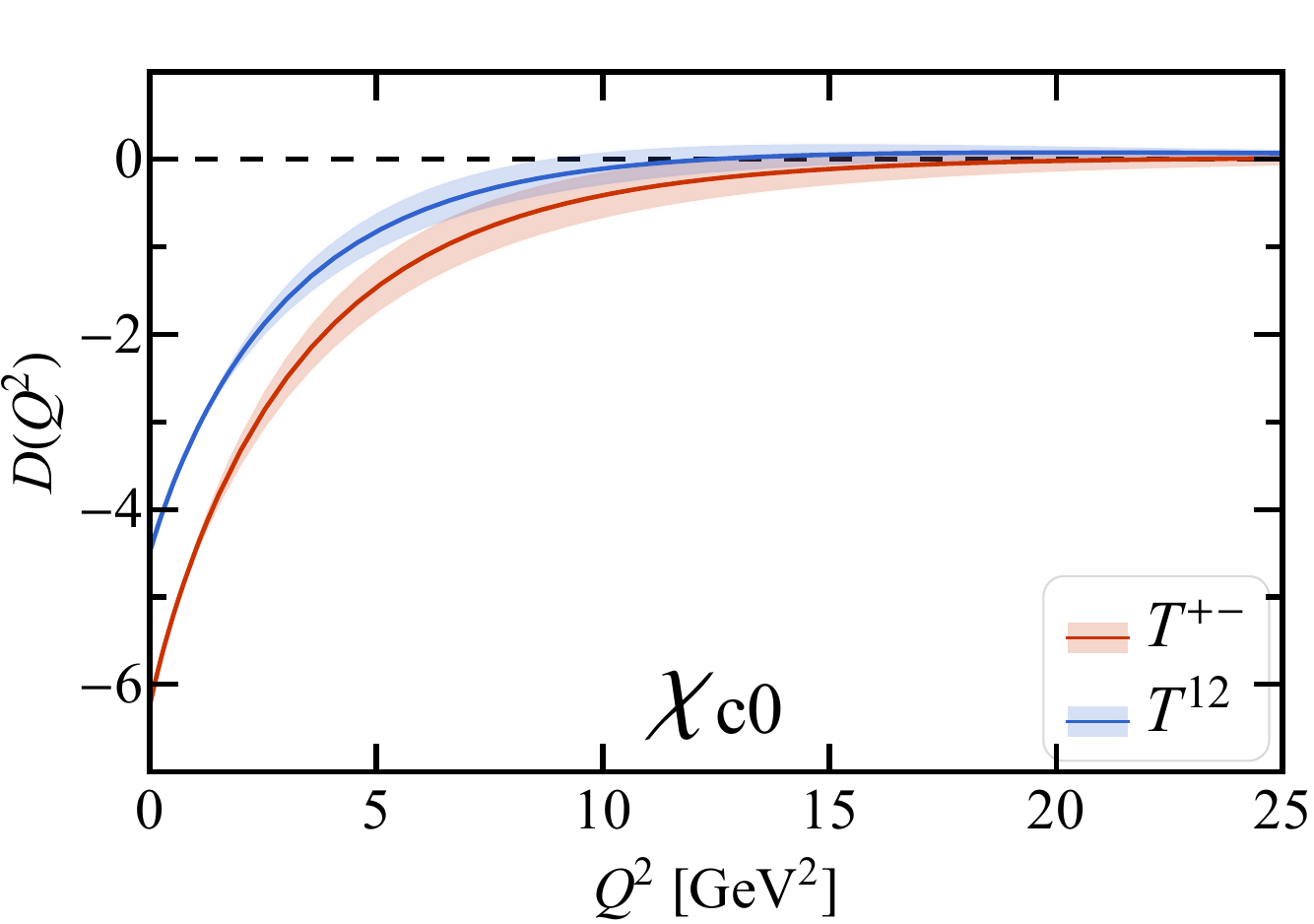}}
\caption{Comparison of the GFF $D(Q^2)$ of $\eta_c$ (\textit{top}) and $\chi_{c0}$ (\textit{bottom}) obtained from the shear $T^{12}$ with our previous result extracted from the light-front energy density $T^{+-}$. We adopt the $N_{\text{max}} = 8$ results as the central values (shown as solid lines), which corresponds to a UV resolution about the same as the hadron mass. We quote the difference between $N_{\max} = 8$ and $N_{\max} = 16$ results to indicate the basis sensitivity (shown as bands).}
\label{fig:charmonium_T12vsT+-}
\end{figure}

Figure~\ref{fig:charmonium_T12vsT+-} compares the GFF $D(Q^2)$ of $\eta_c$ and $\chi_{c0}$ obtained from the shear $T^{12}$ with our previous result extracted from the light-front energy density $T^{+-}$. The obtained $D$-terms for $\eta_c$ are quite close to each other while some discrepancy is observed for $\chi_{c0}$. The comparison suggests that the impulse ansatz is a good approximation for $\eta_c$, and the current is also well conserved,  which also validates our previous work using $T^{+-}$ for $\eta_c$. Note that the impulse ansatz is a better approximation for more compact systems, e.g. $\eta_c$. 
 The prediction of the $D$-term obtained from $T^{12}$ and from $T^{+-}$ as well as the derived radii is collected and compared in Table~\ref{table:D_and_radii} for charmonia $\eta_c$, $\eta_c'$, $\eta_c''$, $\chi_{c0}$ and $\chi_{c0}'$. 
 Here we correct some typographical errors for $T^{+-}$ from Ref.~\cite{Xu:2024cfa}. 
 The results for the $S$-wave pseudoscalars $\eta_c$, $\eta_c'$, $\eta_c''$ show good agreement for the two currents, while the $P$-wave scalars $\chi_{c0}$ and $\chi_{c0}'$ show discrepancies between $T^{12}$ and $T^{+-}$.  

\begin{table}
        \centering
        \renewcommand{\arraystretch}{1.5}
        \caption{BLFQ prediction of the $D$-term $D = D(0)$ extracted from the shear stress $T^{12}$ and the light-front energy density $T^{+-}$ for  charmonia $\eta_c$, $\eta_c'$, $\eta_c''$, $\chi_{c0}$ and $\chi_{c0}'$.  The associated matter radius $r_A$, energy radius $r_E$, invariant mass squared radius $r_{M^2}$ and the scalar radius $r_\theta$ are also obtained for comparison.  We adopt the $N_{\text{max}} = 8$ results as
        the central values, which corresponds to a UV resolution about the same as the hadron mass. We quote the
        difference between $N_{\max} = 8$ and $N_{\max} = 16$ results to indicate the basis sensitivity. The parenthesis provides the change in the last digit. Note that we correct some typographical errors for $T^{+-}$ from Ref.~\cite{Xu:2024cfa}.
        }
        \label{table:D_and_radii}
            \begin{tabular}{@{}ccccccc@{}}
                \toprule
                & & $D$ & $r_A$ (fm) & $r_E$ (fm) & $r_{M^2}$ (fm) & $r_\theta$ (fm)\\
                 \hline
                \multirow{5}{*}{$T^{12}$} & $\eta_c$    & $-4.5(0)$ & 0.18(1) & 0.23(1) & 0.29(1) & 0.33(1) \\
                & $\eta_c'$    & $-6.5(5)$ & 0.35(2) & 0.39(2) & 0.42(2) & 0.45(2) \\
                & $\eta_c''$    & $-12(2)$ & 0.51(4) & 0.55(4) & 0.58(5) & 0.61(5) \\
               &  $\chi_{c0}$ & $-4.5(1)$ & 0.24(2) & 0.27(1) & 0.31(1) & 0.35(1) \\
               &  $\chi_{c0}'$ & $-7.6(8)$ & 0.41(3) & 0.44(3) & 0.47(3) & 0.50(3) \\      
                \hline          
                \multirow{5}{*}{$T^{+-}$} & $\eta_c$    & $-4.7(0)$ & 0.18(1) & 0.23(1) & 0.29(1) & 0.34(0) \\
               &  $\eta_c'$    & $-6.7(4)$ & 0.35(2) & 0.39(2) & 0.42(2) & 0.46(2) \\
               &  $\eta_c''$    & $-12(2)$ & 0.51(4) & 0.54(4) & 0.58(4) & 0.61(4) \\
               &  $\chi_{c0}$ & $-6.2(1)$ & 0.24(2) & 0.29(1) & 0.34(1) & 0.39(1) \\
               &  $\chi_{c0}'$ & $-9.5(6)$ & 0.41(3) & 0.45(3) & 0.49(3) & 0.52(3) \\
                \botrule
            \end{tabular}
    \end{table}

\paragraph{Hadronic matrix element}\label{sect:HME}

Following our work \cite{Cao:2024rul}, the most general covariant decomposition of the hadronic matrix element of the light-front stress-energy tensor for spin-0 particles can be written as, 
\begin{multline}\label{eqn:CLFD_decomposition}
    \langle p' | T_{i}^{\alpha\beta}(0) | p\rangle = 2P^\alpha P^\beta A_i(-q^2) 
    +  2M^2 g^{\alpha\beta}\bar{c}_i(-q^2) \\
    + \frac{1}{2}(q^\alpha q^\beta - q^2 g^{\alpha\beta}) D_i(-q^2) 
    + \frac{M^4 \omega^\alpha \omega^\beta}{(\omega\cdot P)^2}S_{1i}(-q^2) \\
    + (V^\alpha V^\beta + q^\alpha q^\beta) S_{2i}(-q^2),
\end{multline}
where, $P = (p'+p)/2$, $q = p'-p$, $M^2 = p^2 = p'^2$ and $i$ enumerates the constituents, e.g. the quark and the gluon for hadrons. $\omega^\mu = (\omega^+, \omega^-, \vec\omega_\perp) = (0, 2, 0)$ is a null vector indicating the orientation of the quantization surface, the light front $\Sigma = \{x\in \mathbb R^{3,1}|\omega\cdot x = 0\}$. 
Vector $V^\alpha$ is defined as $V^\alpha = \varepsilon^{\alpha \beta \rho \sigma}P_\beta q_\rho \omega_\sigma/(\omega\cdot P)$, which is perpendicular to $q^\alpha$, $P^\alpha$ as well as $\omega^\alpha$. We have adopted the Drell-Yan frame $\omega\cdot q = 0$, which is equivalent to integrating out the light-front longitudinal direction $x^-$ of the operator $T^{\mu\nu}_i(x)$ \cite{Miller:2010nz}. The total stress-energy tensor is obtained by summing over the constituents, viz. $T^{\mu\nu} = \sum_i T^{\mu\nu}_i$. 

Form factors $A_i$, $D_i$ and $\bar c_i$ are the physical GFFs, whereas $S_{1i}$ and $S_{2i}$ are known as spurious GFFs. The spurious GFFs arise because we parametrize the hadronic matrix element with the kinematical subgroup of the Poincaré group in Eq.~(\ref{eqn:CLFD_decomposition}) \cite{Carbonell:1998rj}. Our motivation is that the rest of the Poincaré symmetries are not manifest on the light front, and can be broken in practical calculations in light-front dynamics, especially by Fock space truncation. If full Poincaré symmetry is retained, these spurious GFFs will simply vanish. 
The violation of the Poincaré symmetries may also induce divergences in the GFFs. We have shown in Ref.~\cite{Cao:2024rul} that all uncanceled divergences are contained in the spurious GFFs with decomposition (\ref{eqn:CLFD_decomposition}). 

From Eq.~(\ref{eqn:CLFD_decomposition}), we can express the hadronic matrix elements in terms of the GFFs as follows \cite{Cao:2024rul},
\begin{align}
& t^{++}_i= 2(P^{+})^2 A_i, \label{eqn:t++_CLFD_breit}\\
& t^{+a}_i = 0, \label{eqn:t+a_CLFD_breit} \\
& t^{12}_i = \frac{1}{2} q^1_\perp q^2_\perp  D_i, \label{eqn:t12_CLFD_breit} \\
& t^{+-}_i = 2(M^2+\frac{1}{4}q_\perp^2)A_i  + q^2_\perp D_i  + 4M^2\bar c_i, \label{eqn:t+-_CLFD_breit}  \\
& t^{--}_i = 2 \Big(\frac{M^2+\frac{1}{4}q_\perp^2}{P^+}\Big)^2 A_i + \frac{4M^4}{(P^+)^2}S_{1i}, \label{eqn:t--_CLFD_breit} \\
& t^{-a}_i = 0, \label{eqn:t-a_CLFD_breit} \\
& t^{11}_i + t^{22}_i = - \frac{1}{2} q^2_\perp D_i - 4M^2 \bar c_i + 2q_\perp^2S_{2i}.  \label{eqn:t1122_CLFD_breit} 
\end{align}
Here, $t^{\alpha\beta}_i = \langle p' | T^{\alpha\beta}_i |p\rangle$, and $a = 1,2$ is the transverse spatial index and $i$ enumerates the constituents. The GFFs in the Drell-Yan frame ($q^+=0$) depend on the transverse momentum transfer $q_\perp^2 = -q^2$. Thanks to boost invariants in light-front dynamics, we have further adopted the transverse Breit frame ($\vec P_\perp = (\vec p_\perp + \vec p'_\perp)/2 = 0$, also known as the symmetric frame) to further simplify the expressions.  From these relations, it is straightforward to see that $t^{++}_i$, $t^{12}_i$ and $t^{+-}_i$ do not contain the spurious form factors $S_{1,2}$. Both $t^{+a}_i$ and $t^{-a}_i$ vanish in the Breit frame; however, $t^{+a}_i$ does not contain spurious form factors in the non-Breit frame while $t^{-a}_i$ does. As we mentioned, $t^{++}_i$, $t^{+a}_i$, $t^{12}_i$ and $t^{+-}_i$ are the ``good currents" that can be used to extract the physical GFFs. 

The total GFFs can be obtained by summing over contributions from each constituent. Adopting the good currents, 
\begin{align}
& t^{++} = 2(P^{+})^2 A, \label{eqn:t++_CLFD_breit_total}\\
& t^{12} = \frac{1}{2} q^1_\perp q^2_\perp  D, \label{eqn:t12_CLFD_breit_total} \\
& t^{+-} = 2(M^2+\frac{1}{4}q_\perp^2)A  + q^2_\perp D  + 4M^2\bar c, \label{eqn:t+-_CLFD_breit_total}
\end{align}
where, $A = \sum_i A_i$, $D = \sum_i D_i$, and $\bar c = \sum_i \bar c_i$. In continuum theory, conservation of the stress-energy tensor requires that $\bar c(Q^2) = 0$. In principle, GFF $D$ can be extracted from both $t^{12}$ and $t^{+-}$ (together with $t^{++}$). However, in practical calculations on the light front, $\bar c(Q^2) = 0$ cannot be guaranteed except for the forward limit $Q^2 = 0$. In Ref.~\cite{Cao:2024rul}, we obtained a non-trivial constraint for the good currents $t^{+-}$, $t^{++}$ and $t^{12}$ in order to have a vanishing $\bar c(Q^2)$, 
\begin{equation}
\frac{M^2+\frac{1}{4}\vec q^2_\perp}{(P^+)^2} t^{++} + 2\frac{\vec q^2_\perp}{q^1_\perp q^2_\perp}t^{12} = t^{+-}.
\end{equation}
If this constraint is not fulfilled, we will obtain a non-vanishing total GFF $\bar c(Q^2)$. Since $\bar c_i(Q^2)$ represents the force between the $i$-th constituent and the rest of the system, a non-vanishing $\bar c(Q^2)$ implies that the total force acting on the system is non-vanishing, closely resembling the effect of the cosmological constant $g^{\mu\nu}\Lambda$ \cite{Teryaev:2016edw}. 
Note that, from Eq.~(\ref{eqn:CLFD_decomposition}), current conservation is further violated by the spurious GFF $S_2 = \sum_i S_{2i}$. The difference is that $\bar c(Q^2)$ is free of divergences while $S_2$ may still contain uncanceled divergences \cite{More:2021stk, More:2023pcy}. 

In Ref.~\cite{Xu:2024cfa}, we adopted a slightly different Lorentz decomposition of the hadronic matrix element for the stress-energy tensor and the $\bar c(Q^2)$ was not present in $t^{+-}$. Effectively, we have assumed that the physical part of the stress-energy tensor is conserved, and the possible non-vanishing $\bar c(Q^2)$ is absorbed in the GFF $D(Q^2)$.  This method does not affect the von Laue condition $\lim_{Q^2\to 0}Q^2D(Q^2) = 0$ since $\bar c(0) = 0$, and the system is still in mechanical equilibrium. However, the obtained GFF $D$ will be different from the $D$-term extracted from $T^{12}$, as shown in Table~\ref{table:D_and_radii}. Fortunately, Fig.~\ref{fig:etac_1S_T12vsT+-} shows that the contribution arising from $\bar c(Q^2)$ is very small for the ground-state charmonium $\eta_{c}$. 

\paragraph{Basis light front quantization}\label{sect:BLFQ}

In Hamiltonian formalism, the hadronic state vector can be solved from the Hamiltonian Schrödinger equation \cite{Namyslowski:1985zq, Zhang:1994ti, Burkardt:1995ct, Brodsky:1997de, Carbonell:1998rj, Heinzl:2000ht, Miller:2000kv, Bakker:2013cea, Hiller:2016itl, Brodsky:2022fqy}, 
\begin{equation}\label{eqn:HEV}
H_\textsc{lc} |p\rangle = M^2|p\rangle, 
\end{equation}
where, the light-cone Hamiltonian $H_\textsc{lc}$ is defined as $H_\textsc{lc}  = P^+ P^- - \vec P_\perp^2$. 
In Ref.~\cite{Xu:2024cfa}, we adopt the basis light-front quantization (BLFQ) approach to solve the charmonium system in light front QCD \cite{Vary:2009gt}. 
Specifically, we adopted a low-energy effective Hamiltonian for charmonium, which only involves the valence Fock sectors $|c\bar c\rangle$ \cite{Li:2015zda, Li:2017mlw}. The effective interaction incorporates the soft-wall AdS/QCD confining potential along with a longitudinal confining potential $V_\text{conf}$ as well as a one-gluon exchange interaction $V_\text{OGE}$ derived from light-front QCD \cite{Li:2017mlw}, 
\begin{equation}
H_\text{eff} = T + V_\text{conf} + V_\text{OGE},
\end{equation}
where, $T = \sum_i (k^2_{i\perp}+m_i^2)/x_i$ is the light-front kinetic energy. 

To solve the Hamiltonian eigenvalue problem (\ref{eqn:HEV}), we represent the Hamiltonian as a matrix $\big[H \big]_{ij} = \langle i | H | j \rangle$ within a prescribed basis $|i\rangle$. The basis was chosen as the harmonic oscillator functions in the transverse direction and the Jacobi polynomials in the longitudinal direction \cite{Li:2015zda}. This basis preserves all kinematical symmetries of the QCD Hamiltonian and is also the eigenbasis of the AdS/QCD Hamiltonian $H_0 = T + V_\text{conf}$ \cite{Li:2015zda}. 
The Hamiltonian matrix is then numerically diagonalized to obtain the mass spectrum as well as the wave functions, which are now available on Mendeley Data \cite{Li:2019}. The charmonium spectrum is shown to be in good agreement with the experimental measurements \cite{Li:2017mlw}. The obtained light-front wave functions are used to compute a variety of observables including the radiative widths and provide good agreement with the experimental values \cite{Li:2017mlw, Li:2018uif, Adhikari:2018umb, Li:2021ejv, Wang:2023nhb}. 

From the valence sector ($c\bar c$) wave functions $\psi_{s\bar s}(x, \vec k_\perp)$, the state vector can be represented as, 
\begin{multline}
|p\rangle = \sum_{s, \bar s} \int_0^1 \frac{\dd x}{2x(1-x)} \int \frac{\dd^2k_\perp}{(2\pi)^3} \\
\times \psi_{s\bar s}(x, \vec k_\perp)  \frac{1}{\sqrt{N_c}} \sum_{i=1}^{N_c} b^\dagger_{si}\big(xp^+, \vec k_\perp+x\vec p_\perp\big) \\
\times d^\dagger_{\bar si}\big((1-x)p^+, -\vec k_\perp+(1-x)\vec p_\perp\big) |0\rangle.
\end{multline}
Here, $N_c = 3$ and $p^\mu = (p^-, p^+, \vec p_\perp)$ is the 4-momentum of the bound state. 
$\vec k_\perp = \vec p_{c\perp} - x\vec p_\perp$ is the relative transverse momentum of the $c$-quark, and $x = p^+_c / p^+$ is its longitudinal momentum fraction. $b^\dagger$ and $d^\dagger$ are the creation operators for the quark and antiquark, respectively. They are related to the quark field operator $\psi$ as, 
\begin{equation}
\psi(0) = \sum_s \int \frac{\dd^3p}{(2\pi)^32p^+} \Big\{b_s(p) u_s(p) + d^\dagger_s(p) v_s(p) \Big\},
\end{equation} 
where, $u, v$ are the covariant 4-component Dirac spinors. 

The QCD stress-energy tensor consists of the quark part and the gluon part as well as their interactions. Since our charmonium state defined at low-energy resolution contains just the effective quark and antiquark degree of freedom, only the quark part of the stress-energy tensor is active. Note that the obtained GFFs are still the total GFFs, not the quark GFFs (cf. \cite{Nair:2024fit}). The reason is as follows. The total GFFs are scale independent, i.e. they can be evaluated at any scale. At the low-energy resolution like our model, only the effective quark degree of freedom is active and the total GFFs evaluated at this scale is the same as the effective quark GFFs. The wave function representation of the shear stress is \cite{Cao:2024fto}, 
\begin{multline}\label{eqn:t12}
t^{12} = \frac{1}{2} \sum_{s, \bar s} \int \frac{\dd x}{4\pi x(1-x)}\int \dd^2 r_\perp \widetilde\psi_{s\bar s}^*(x, \vec r_\perp) \\
\times \sum_{j=c,\bar c} e^{-i\vec q_\perp \cdot \vec r_{j\perp}} \frac{i\tensor\nabla^1_{j\perp}i\tensor\nabla^2_{j\perp}-q^1_\perp q^2_\perp}{x_j}
\widetilde\psi_{s\bar s}(x, \vec r_\perp). 
\end{multline}
This expression can be extended to the higher Fock sectors \cite{Cao:2024fto}. Gluons are incorporated by replacing the canonical momentum with mechanical momentum. 
A similar expression in the non-relativistic limit is derived by Freese \cite{Freese:2024rkr}. From the above expression, the $D$-term is closely related to the classical concept virial $\vec r\cdot \vec p$, which measures the dilation behavior of the system \cite{Lorce:2021xku, Cao:2023ohj, Lorce:2025oot}.

\paragraph{Numerical results}\label{sect:numerics}

In practical calculations, the BLFQ results come with a basis truncation labelled $N_{\mathrm{max}}$, which corresponds to a UV resolution $\Lambda_\textsc{uv} \sim \kappa\sqrt{N_\text{max}}$, where $\kappa \approx 1\,\mathrm{GeV}$ is the basis scale parameter of the harmonic oscillator wave function \cite{Li:2017mlw}. Following our previous practice \cite{Li:2017mlw, Li:2021ejv, Wang:2023nhb}, we adopt the $N_{\text{max}} = 8$ results as the central value, which corresponds to a UV resolution about the same as the charmonium mass $\Lambda_\textsc{uv} \approx M$. We adopt the difference between $N_{\max} = 8$ and $N_{\max} = 16$ results to indicate the basis sensitivity. 

\begin{figure}
\centering
\subfigure[\ \label{fig:etac_nS_T12_D}]{\includegraphics[width=0.4\textwidth]{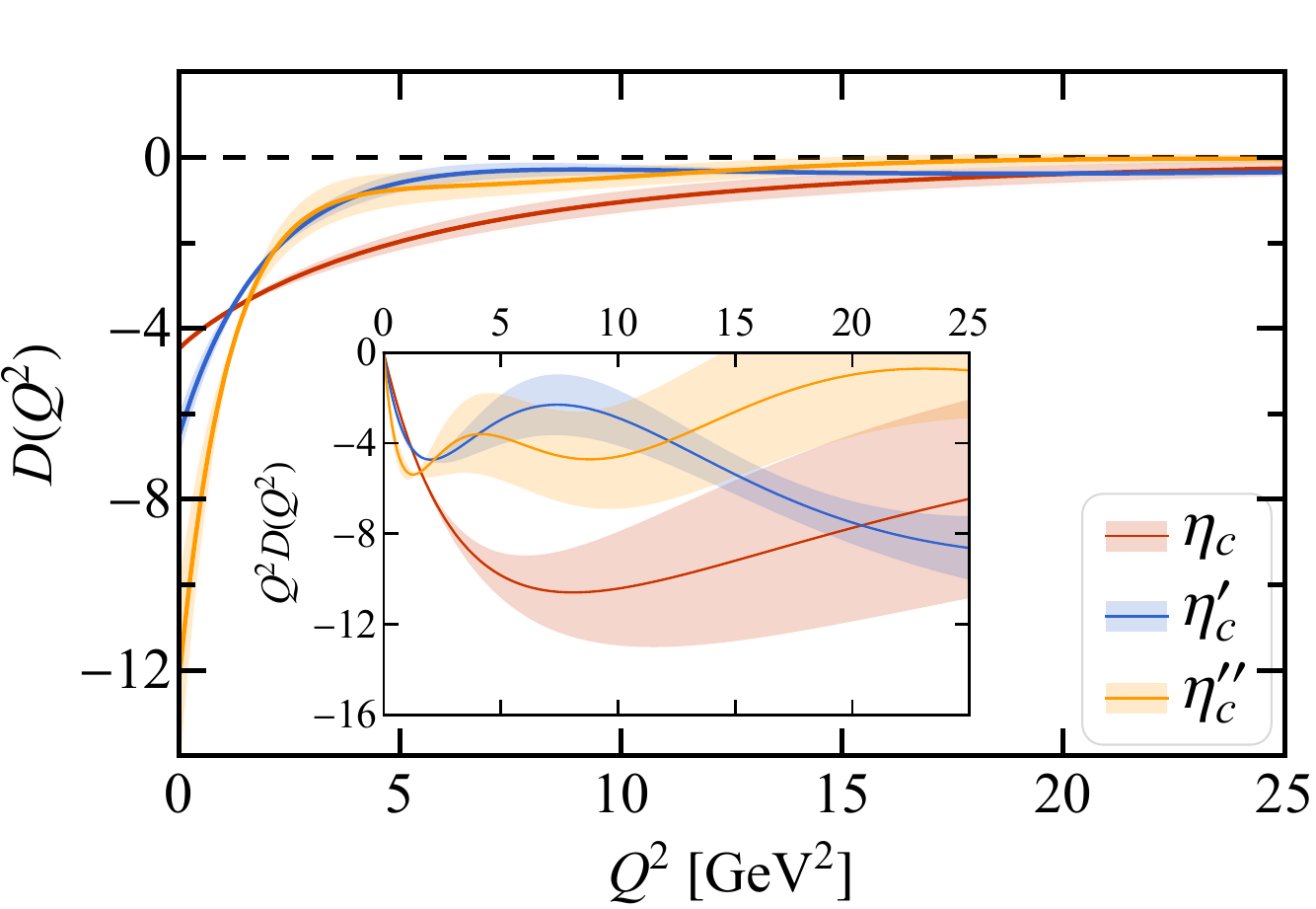}}
\subfigure[\ \label{fig:etac_nS_T12_C}]{\includegraphics[width=0.41\textwidth]{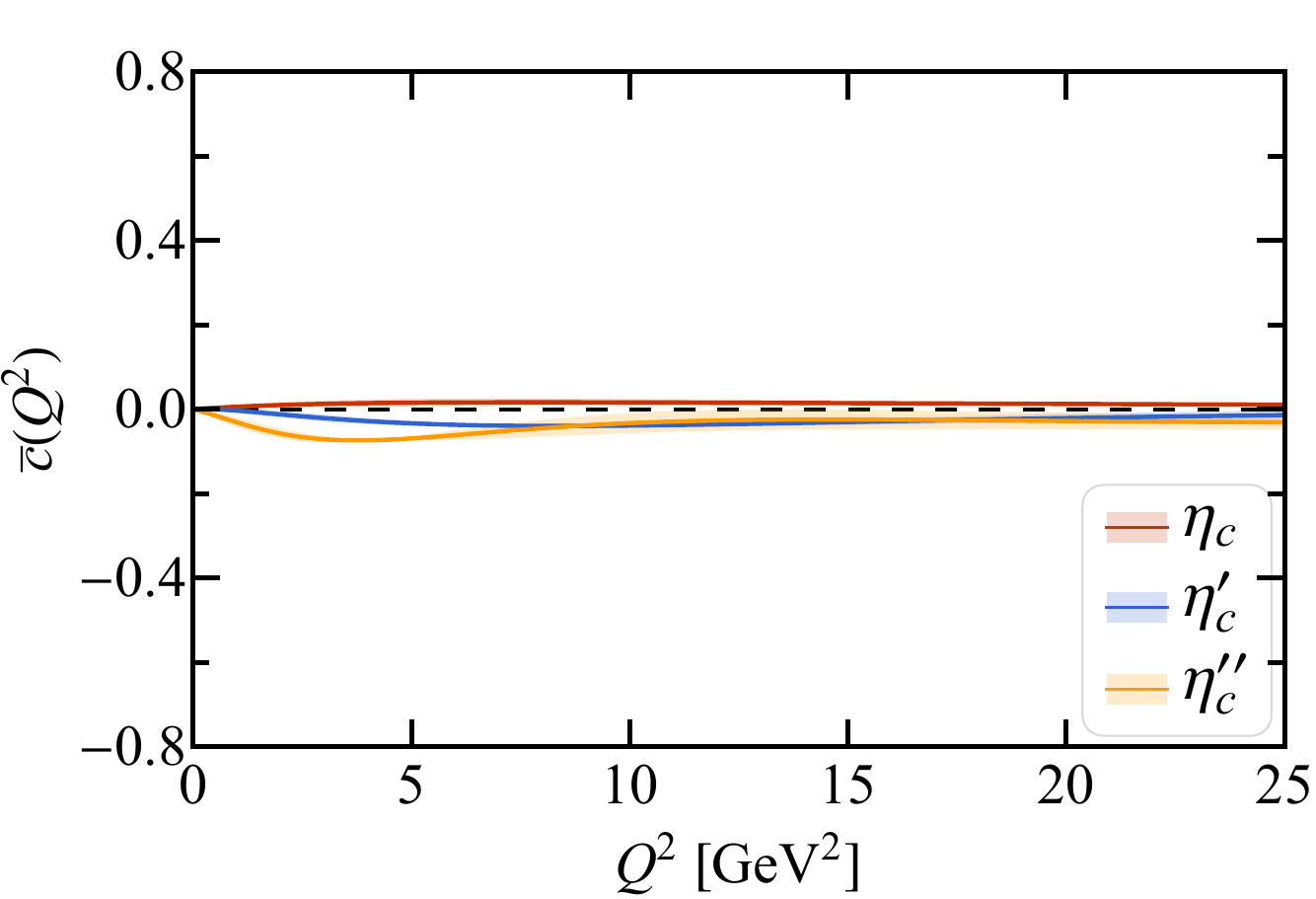}}
\caption{(\textit{Top}) GFF $D(Q^2)$ of charmonium $\eta_c$ and its radial excitations $\eta'_c$ and $\eta''_c$ obtained from the shear stress $T^{12}$. 
(\textit{Bottom}) GFF $\bar c(Q^2)$ of charmonia $\eta_c$, $\eta'_c$ and $\eta''_c$ obtained by combining results from shear stress $T^{12}$ and from the light-front energy density $T^{+-}$. In the continuum limit, $\bar c(Q^2)$ vanishes due to the conservation of the stress-energy tensor. 
We adopt the $N_{\text{max}} = 8$ results as the central values and the difference between $N_{\max} = 8$ and $N_{\max} = 16$ results as the bands. See the caption of Fig.~\ref{fig:charmonium_T12vsT+-} for more details.}
\label{fig:etac_nS_T12}
\end{figure}
  
\begin{figure}
\centering
\,\,\,\,\,\subfigure[\ \label{fig:chic0_nP_T12_D}]{\includegraphics[width=0.4\textwidth]{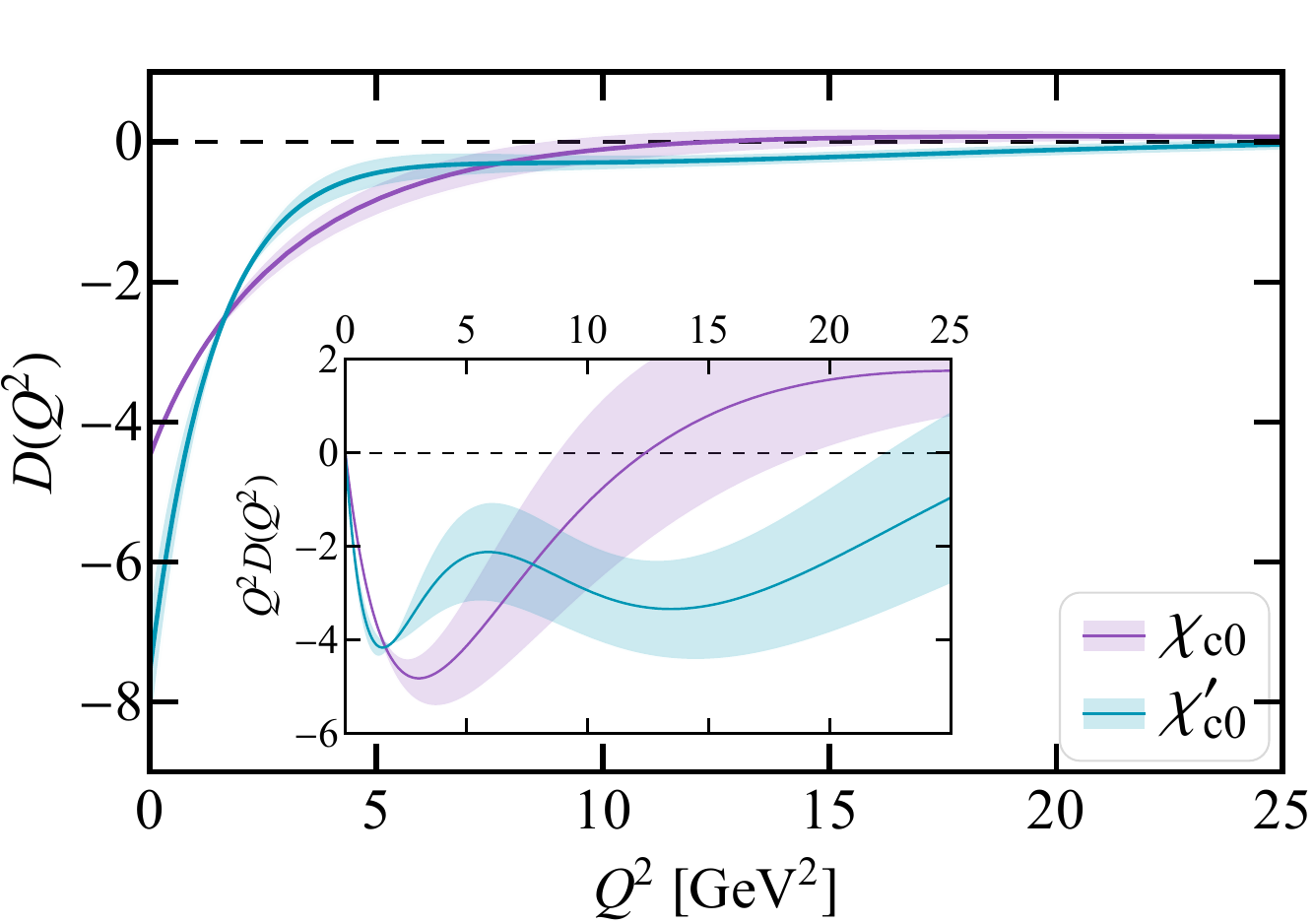}}
\subfigure[\ \label{fig:chic0_nP_T12_C}]{\includegraphics[width=0.42\textwidth]{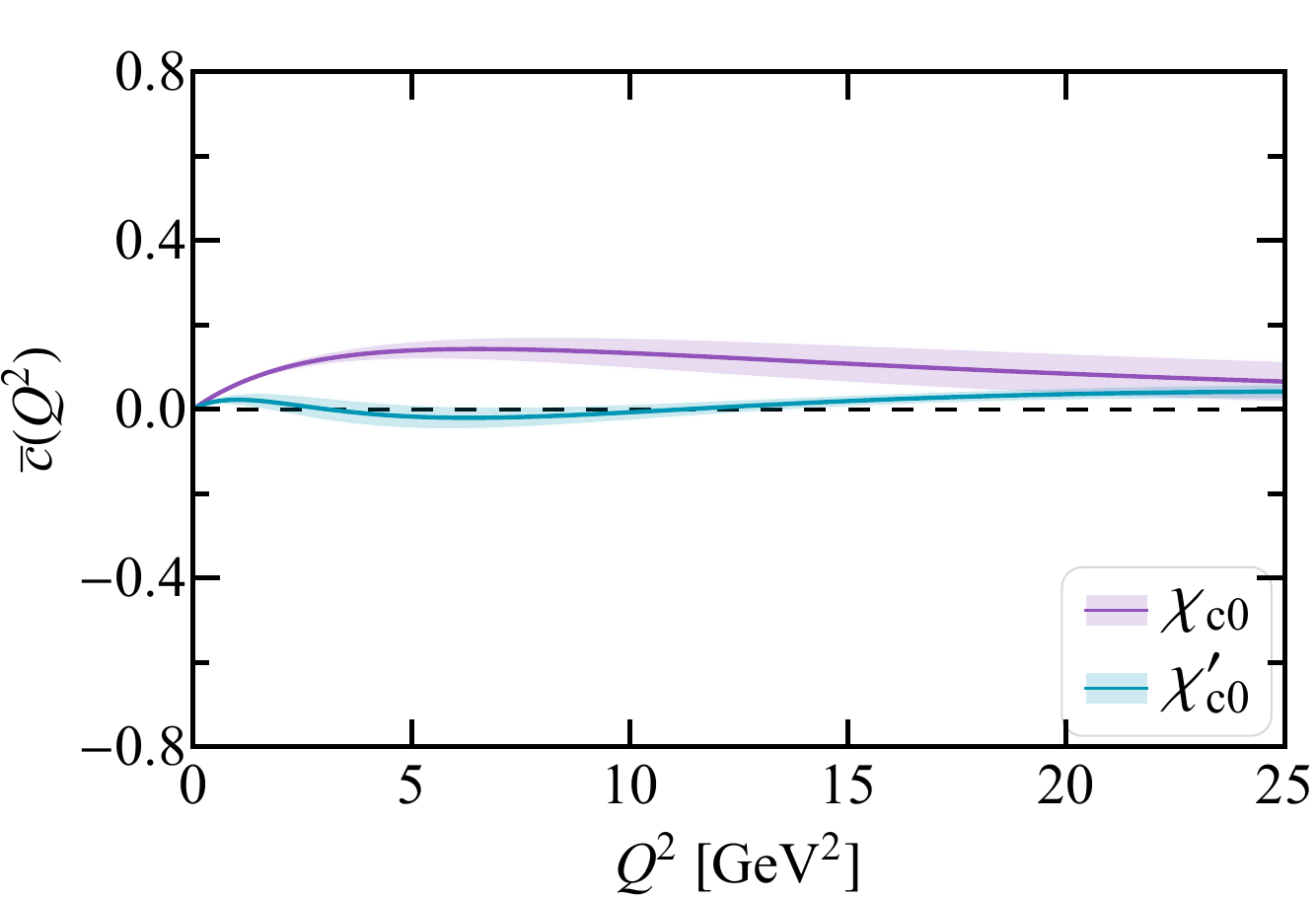}}
\caption{The same as Fig.~\ref{fig:etac_nS_T12}, except for $P$-wave charmonia $\chi_{c0}$ and $\chi'_{c0}$. }
\label{fig:chic0_nP_T12}
\end{figure}

Figure~\ref{fig:etac_nS_T12_D} shows the GFF $D(Q^2)$ of the S-wave charmonia $\eta_c$, $\eta'_c$ and $\eta''_c$ extracted from the shear stress $T^{12}$. The $D$-term of each particle is finite and increases at small $Q^2$ as the nodes of the radial excitation increases. The GFF $D(Q^2)$ also exhibits oscillations for radially excited states, which is better shown on the $Q^2D(Q^2)$ vs $Q^2$ inset. By combining these results with our previous results using $T^{+-}$, we can extract the GFF $\bar c(Q^2)$. As mentioned, this form factor is expected to vanish to conserve the stress-energy tensor current. Thus, we can use this form factor to gauge the violation of the current conservation. Fig.~\ref{fig:etac_nS_T12_C} shows the GFF $\bar c(Q^2)$ of the S-wave charmonia $\eta_c$, $\eta'_c$ and $\eta''_c$. In the forward limit $Q^2 = 0$, this form factor vanishes for all states as expected. It remains a small but non-vanishing value in the off-forward limit. 
The GFF $D(Q^2)$ of the $P$-wave charmonia $\chi_{c0}$ and $\chi'_{c0}$ are shown in Fig.~\ref{fig:chic0_nP_T12_D}. Similarly, the current conservation violating GFF $\bar c(Q^2)$ of the $P$-wave charmonia is shown in Fig.~\ref{fig:chic0_nP_T12_C}. 
Note that the radially excited states, e.g. $\eta'_c, \eta''_c, \chi'_{c0}$, all develop nodal structures, resulting from the nodal structures of their densities according to Eq.~(\ref{eqn:t12}).
The obtained $D$-terms of the ground states $\eta_c$ and $\chi_{c0}$ are of the same order of magnitude for the proton \cite{Hackett:2023rif, Burkert:2018bqq, Duran:2022xag}, despite the fact that the former is a more compact system, suggesting similar scaling behaviors of the strong force within these two hadronic systems.

\begin{figure}
\centering
\subfigure[\ \label{fig:chic0_nP_T12_PEM}]{\includegraphics[width=0.4\textwidth]{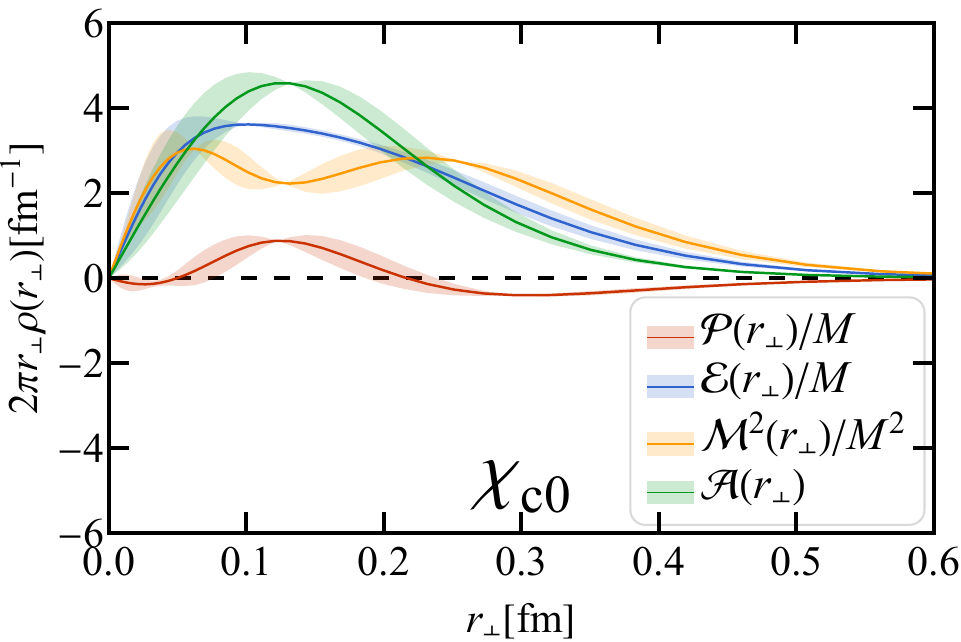}}
\subfigure[\ \label{fig:chic0_nP_T12vsT+-_pressure}]{\includegraphics[width=0.415\textwidth]{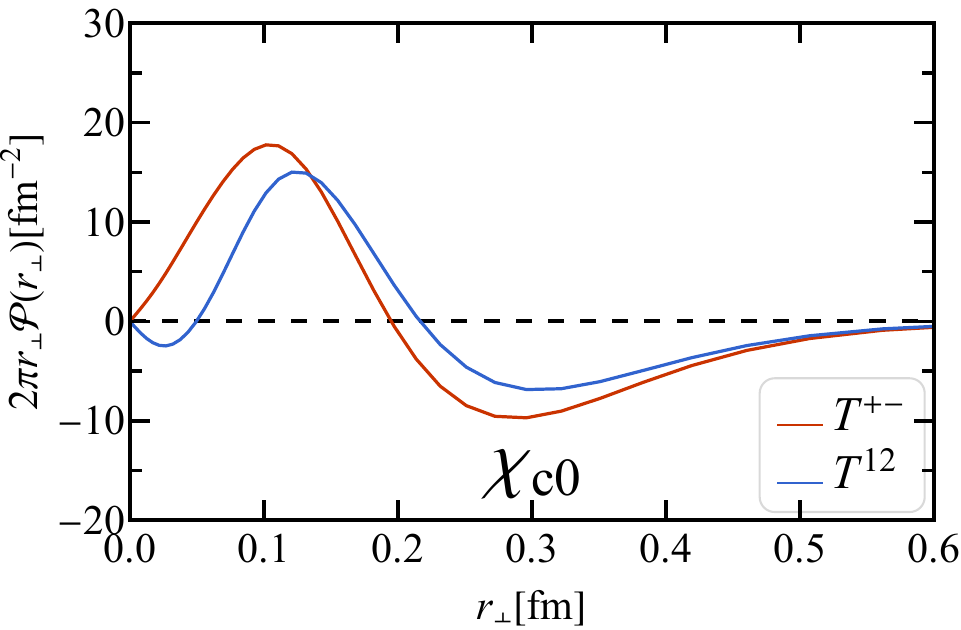}}
\caption{
(\textit{Top}) Transverse densities of $\chi_{c0}$: pressure $\mathcal P$, energy density $\mathcal E$, invariant mass squared density $\mathcal M^2$ and matter density $\mathcal A$ as extracted from $T^{12}$.
We adopt the $N_{\text{max}} = 8$ results as the central values and the difference between $N_{\max} = 8$ and $N_{\max} = 16$ results as the bands. See the caption of Fig.~\ref{fig:charmonium_T12vsT+-} for more details. (\textit{Bottom}) Comparison of the pressure of $\chi_{c0}$ obtained from $T^{+-}$ and $T^{12}$ with $N_{\mathrm{max}} = 8$. Note that the pressure extracted from $T^{12}$ exhibits an attractive core, in contrast to the argument of a repulsive core based on mechanical stability \cite{Mai:2012cx, Mai:2012yc, Cantara:2015sna, Hudson:2017oul, Polyakov:2018zvc}. 
}
\label{fig:chic0_densities}
\end{figure}

From the GFFs $A$ and $D$, we can compute a variety of transverse densities of the system, including the matter density $\mathcal A(r_\perp)$, energy density $\mathcal E(r_\perp)$, pressure $\mathcal P(r_\perp)$, invariant mass squared density $\mathcal M^2(r_\perp)$ and the scalar density $\theta(r_\perp)$. The precise meaning and definition of these densities can be found in our previous work \cite{Xu:2024cfa, Maynard:2024wyi}. Fig.~\ref{fig:chic0_nP_T12_PEM} shows the relevant transverse densities of $\chi_{c0}$, whose GFF $D(Q^2)$ show noticeable difference from our previous work using $T^{+-}$ (see Fig.~\ref{fig:chic0_1P_T12vsT+-}). The pressure is particularly intriguing. Pressure of $\chi_{c0}$ extracted from $T^{12}$ exhibits an attractive core, as shown in Fig.~\ref{fig:chic0_nP_T12vsT+-_pressure}. However, it still satisfies the von Laue condition and with a negative $D(0)$. Therefore, it provides a counter example to the speculation that a mechanically stable system should have a repulsive core with an attractive periphery \cite{Mai:2012cx, Mai:2012yc, Cantara:2015sna, Hudson:2017oul, Polyakov:2018zvc}. Note that pressure extracted from $T^{+-}$ is still with a repulsive core. The attractive core from $T^{12}$ is associated with the zero crossing of the GFF $D(Q^2)$ at large $Q^2$ as shown in Fig.~\ref{fig:chic0_1P_T12vsT+-}.

\paragraph{Summary} 

In this work, we investigate the charmonium gravitational form factor $D$ extracted from the shear stress $T^{12}$, and compare the results with our previous results obtained from the light-front energy density $T^{+-}$. The non-perturbative wave function is obtained from basis light-front quantization. Our analysis is based on the recent covariant light-front analysis of the hadronic matrix element of the stress-energy tensor, which removes uncanceled divergences in the physical gravitational form factors. From this analysis, both $T^{+-}$ and $T^{12}$ are good currents that can be used to extract GFF $D$. The latter has several advantages: first, it does not require one to make an assumption of current conservation; second, it does not rely on the impulse ansatz. On the other hand, we have found (but have not shown some of the comparisons here) that GFF $D$ extracted from these two currents are close to each other for several of the low-lying charmonia, e.g. $\eta_c$, $\eta'_c$ and $\chi'_{c0}$. Indeed, by combining $T^{12}$ and $T^{+-}$, we are able to show that there is a small violation of the current conservation in our formalism. 

The obtained charmonium GFFs as well as their microscopic wave function representation may be helpful to understand the properties of the hadronic stress-energy as well as the internal structures of QCD bound states. 
The method we have developed can be extended to other hadrons, such as the nucleons and the pion, which have been solved in the light-front Hamiltonian formalism \cite{Honkanen:2010rc, Zhao:2014xaa, Wiecki:2014ola, Hu:2020arv, Nair:2022evk, Nair:2023lir, Li:2021jqb, Li:2022izo, Li:2022ytx, Li:2023izn, Li:2015zda, Li:2017mlw, Leitao:2017esb, Li:2018uif, Adhikari:2018umb, Tang:2018myz, Lan:2019img, Tang:2019gvn, Tang:2020org, Li:2021ejv, Wang:2023nhb, Jia:2018ary, Lan:2019vui, Lan:2019rba, Qian:2020utg, Mondal:2021czk, Adhikari:2021jrh, Li:2022mlg, Zhu:2023lst, Mondal:2019jdg, Xu:2021wwj, Liu:2022fvl, Hu:2022ctr, Peng:2022lte, Kuang:2022vdy, Lan:2021wok, Xu:2022abw, Zhu:2023lst, Zhang:2023xfe, Zhu:2023nhl, Xu:2023nqv, Lin:2023ezw, Kaur:2024iwn, Liu:2024umn, Yu:2024mxo, Lan:2024ais}. 
An interesting question emerging from this work is how to restore current conservation. Analysis along the line of covariant light front dynamics seems to be quite promising. 

\section*{Acknowledgements}

 Y.L. acknowledges support from the National Natural Science Foundation of China (NSFC) under Grant No.~12375081, and from the Chinese Academy of Sciences under Grant No.~YSBR-101.  
%
X.Z. is supported by new faculty startup funding by the Institute of Modern Physics, Chinese Academy of Sciences, by Key Research Program of Frontier Sciences, Chinese Academy of Sciences, Grant No.~ZDBS-LY-7020, by the Foundation for Key Talents of Gansu Province, by the Central Funds Guiding the Local Science and Technology Development of Gansu Province, Grant No.~22ZY1QA006, by Gansu International Collaboration and Talents Recruitment Base of Particle Physics (2023-2027), by International Partnership Program of the Chinese Academy of Sciences, Grant No.~016GJHZ2022103FN, by National Natural Science Foundation of China, Grant No.~12375143, by National Key R\&D Program of China, Grant No.~2023YFA1606903 and by the Strategic Priority Research Program of the Chinese Academy of Sciences, Grant No.~XDB34000000, by the Senior Scientist Program funded by Gansu Province, Grant No. 25RCKA008.
The numerical calculations in this paper have been partly done on the supercomputing system in the Dongjiang Yuan Intelligent Computing Center.
 

\end{document}